\documentclass[aps,prl,reprint,superscriptaddress,nobibnotes,showpacs]{revtex4-1}

\usepackage{bbm}
\usepackage{graphicx}
\usepackage{dcolumn}
\usepackage{bm}
\usepackage[sans]{dsfont}

\newcommand{\ket}[1]{|#1\rangle}
\newcommand{\braket}[2]{\langle{#1}|{#2}\rangle}

\newcommand{\fig}[1]{figure~\ref{#1}}

\usepackage{hyperref}

\def\beq{\begin{eqnarray}}
\def\eeq{\end{eqnarray}}
\def\bfig{\begin{figure}}
\def\efig{\end{figure}}
\usepackage{placeins}

\begin{document}
\title{Comment on ``Non-monotonic projection probabilities as a function of distinguishability''}

\author{Young-Sik Ra}
\affiliation{Department of Physics, Pohang University of Science and Technology (POSTECH), Pohang, 790-784, Korea}

\author{Malte C. Tichy}
\affiliation{Department of Physics and Astronomy, University of Aarhus, DK--8000 Aarhus C, Denmark}

\author{Hyang-Tag Lim}
\affiliation{Department of Physics, Pohang University of Science and Technology (POSTECH), Pohang, 790-784, Korea}

\author{Osung Kwon}
\affiliation{Department of Physics, Pohang University of Science and Technology (POSTECH), Pohang, 790-784, Korea}

\author{Florian Mintert}
\affiliation{Freiburg Institute for Advanced Studies, Albert-Ludwigs-Universit\"at, Albertstrasse 19, D-79104 Freiburg, Germany}
\affiliation{Department of Physics, Imperial College London, London SW7 2AZ, United Kingdom}

\author{Andreas Buchleitner}
\affiliation{Physikalisches Institut der Albert-Ludwigs-Universit\"at, Hermann-Herder-Str.~3, D-79104 Freiburg, Germany}

\author{Yoon-Ho Kim}
\email{Email: yoonho72@gmail.com}
\affiliation{Department of Physics, Pohang University of Science and Technology (POSTECH), Pohang, 790-784, Korea}

\date{\today}

\begin{abstract}A recent work (2014 \textit{New J. Phys.} \textbf{16} 013006) claims that 
nonmonotonic structures found in the many-particle quantum-to-classical transition 
(2013 \textit{Proc. Natl Acad. Sci. USA}
\textbf{110} 1227--1231; 2011 \textit{Phys. Rev. A} \textbf{83} 062111) are not exclusive to the many-body domain, but they also appear for single-photon as well as for semi-classical systems. We show that these situations, however, do not incorporate any quantum-to-classical transition, which makes the claims unsustainable.
\end{abstract}


\maketitle
Recently, reference~\cite{bjork} reported on the nonmonotonic dependence of projection probabilities of quantum states, observed for a single-photon state and semiclassical states. Based on this observation, reference~\cite{bjork} argues that the nonmonotonic change of many-particle event probabilities observed in references~\cite{ra,tichy} is ``not unnatural" and ``not a manifestation of a quantum-to-classical transition". Here, we show that the situation considered in reference~\cite{bjork} is qualitatively different from the scenario in references~\cite{ra,tichy}, such that the conclusions of reference~\cite{bjork} cannot be sustained.

\begin{figure}[b]
\raggedleft{\includegraphics[width=\linewidth]{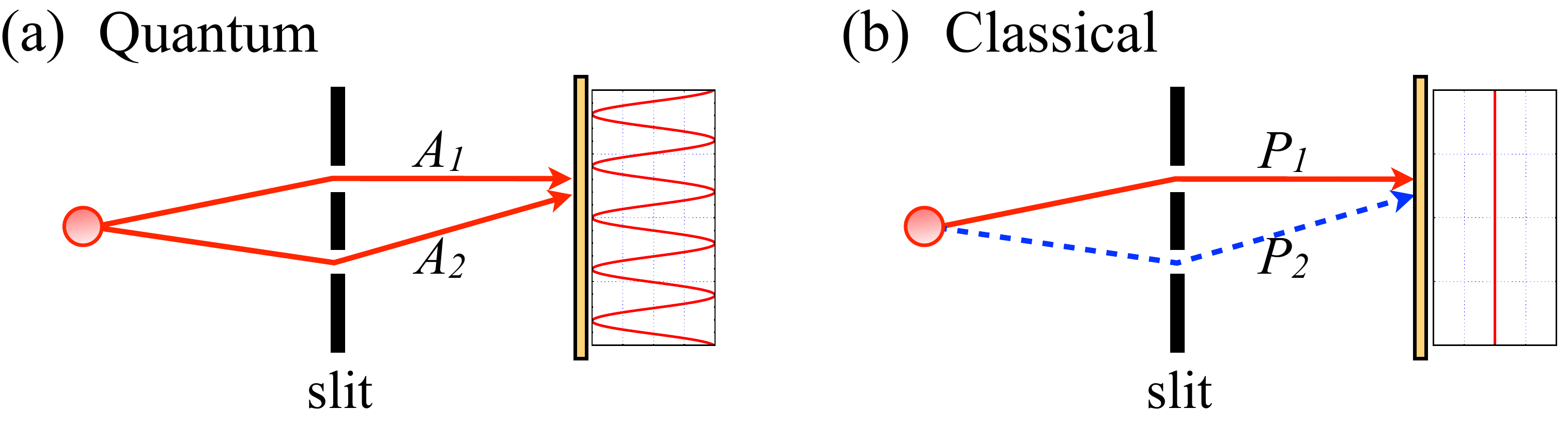}}
\caption{
The double-slit experiment with a single particle. In (a), interference is observed, which requires a quantum description: \emph{probability amplitudes} $A_1$ and $A_2$ are added. On the contrary, in (b), no interference is observed, and a classical description is sufficient: \textit{probabilities} $P_1$ and $P_2$ are added.
}
\label{fig:sketch}
\end{figure}

Let us first rigorously establish the quantum-to-classical transition considered in references~\cite{ra,tichy}.
In the quantum realm, event probabilities are determined by the sum of complex \emph{probability amplitudes} of all processes leading to the event in question, see figure~\ref{fig:sketch}(a): the probability amplitudes are added, and the absolute value of the sum is squared. In the classical realm, event probabilities are obtained by adding the \emph{probabilities} of all processes leading to the event, see figure~\ref{fig:sketch}(b). 
Within the quantum regime, due to the interference of probability amplitudes, event probabilities oscillate as a function of the relative phases between the amplitudes.
Such phase-dependence is absent in the classical realm.

\begin{figure}[t]
\raggedleft{\includegraphics[width=\linewidth]{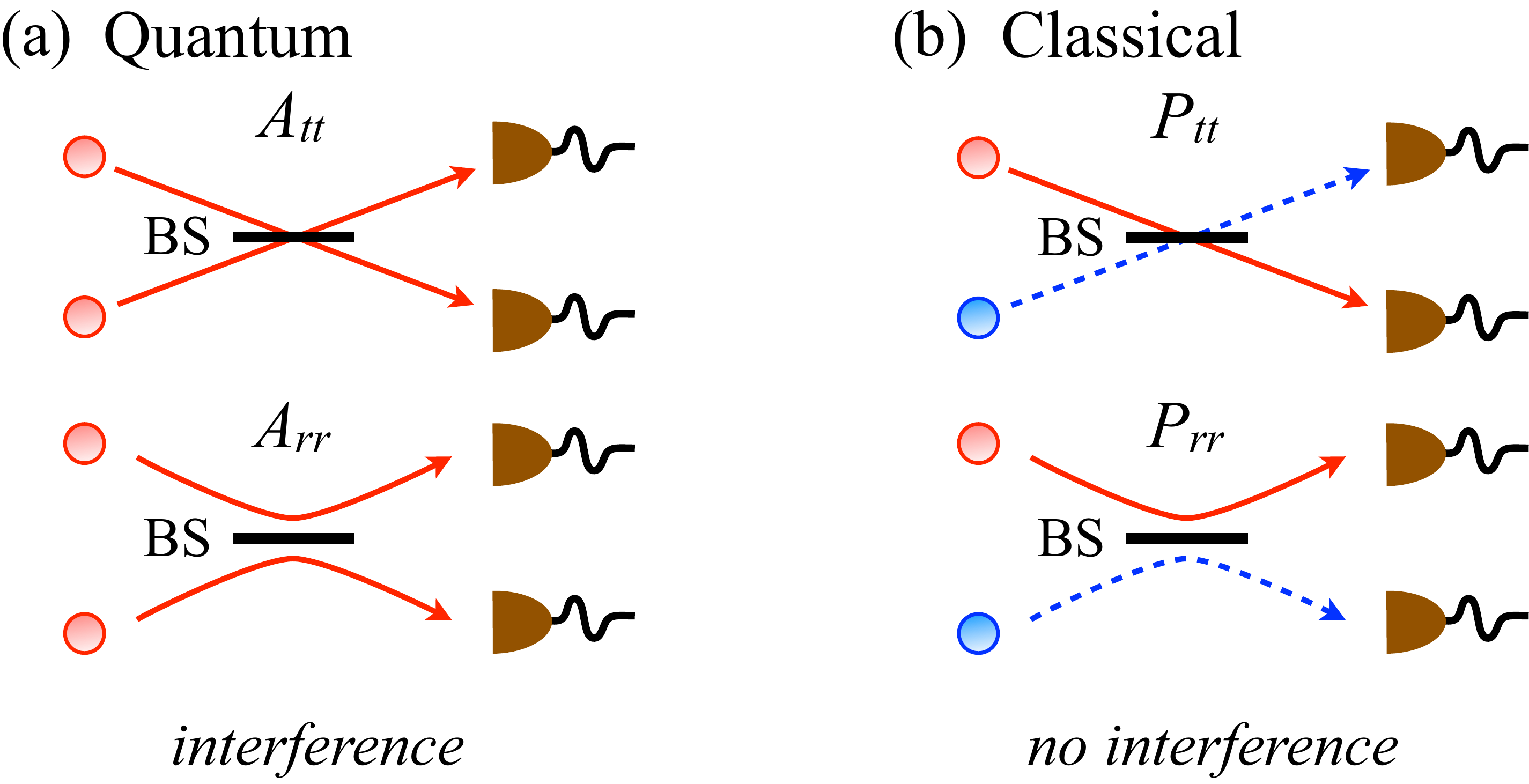}}
\caption{
The Hong--Ou--Mandel experiment using a 50:50 beamsplitter (BS). Two-particle paths of both photons transmitted (upper panels) and both photons reflected (lower panels) (a) interfere in the quantum realm, but (b) they do not in the classical realm. 
Therefore, probability amplitudes $A_{tt}$ and $A_{rr}$ are added in the former, but probabilities $P_{tt}$ and $P_{rr}$ are added in the latter.
}
\label{fig:sketch2}
\end{figure}

The situation sketched in figure~\ref{fig:sketch} applies to, both, single- and many-particle interference. For example, in the two-particle experiment sketched in \fig{fig:sketch2}, \emph{two-particle} amplitudes ($A_{tt}$ and $A_{rr}$) of two indistinguishable particles need to be added to infer the quantum mechanical coincidence event probability, while the sum of the probabilities $P_{tt}$ and $P_{rr}$ matches the event probability in the classical realm. Similar to single-particle interference, many-particle interference can also exhibit phase-dependent oscillations when the relative phases between many-particle amplitudes are varied~\cite{tichy,TutorialMalte,Laloe}.

The transition to the classical regime is induced by lifting the indistinguishability of the processes that contribute to an event~\cite{Feynman, arndt1, arndt2}.
In particular, the detection of any distinctive property of such process (whether controlled by the experimentalist or uncontrolled by the environment) provides which-way information, hence renders different processes distinguishable, and leads to a reduced (if not completely suppressed) ability to interfere~\cite{englert}. 
 For a single particle that can take two slits to fall onto a screen, the observation of the path of the particle leads to the breakdown of the interference pattern; in close analogy, the path delay between identical particles controls the indistinguishability of the interfering processes (i.e.~of the many-particle paths) \cite{hong}. We emphasize that, during such transition, acquiring which-path information does not change the amplitude of either path, but it solely affects the degree to which probabilities, instead of probability amplitudes, are added.

Intuitively, one may expect that the total event probability  can always be written as 
\beq \mathcal{P}= (1-\alpha) P_{\textrm{classical}} + \alpha P_{\textrm{quantum}},\label{intuitiveexp} \eeq
 where $\alpha$ $\in[0;1]$ suitably quantifies the degree of interference and mediates the transition from the event probability in quantum realm $P_{\textrm{quantum}}$ to that in classical realm $P_{\textrm{classical}}$. Any dependence on the relative phase or other parameters is implicit in $P_{\textrm{quantum}}$. This relation remains valid for one- or two-particle interference, in which only fully quantum and fully classical terms arise. Therefore, every observable that is monitored during the transition between the quantum and classical behavior of one or two particles features a monotonic dependence on the degree of coherence $\alpha$. However, references~\cite{ra,tichy} report that multi-particle event probabilities can be nonmonotonic, which rules out a generalization of the intuitive description inherent to equation~(\ref{intuitiveexp}). Instead of a purely classical and a purely quantum contribution, the probability for an $N$-particle event needs to be written as 
  \beq \label{manypevent} \mathcal{P}=  \sum_{d=0}^{N/2}  W_d(\alpha) P_{d},\eeq where $d$ is the number of particles that interfere, and $W_d(\alpha)$ and $P_d$ are the corresponding weight and the associated event probability, respectively. The case $d=0$ can be identified with the classical, $d=N$ with the quantum realm. Since the probabilities $P_d$ do not obey any ordering (in general, neither $P_0 \le P_1 \le P_2 \dots $, nor $P_0 \ge P_1 \ge P_2 \dots $), the event probability described by equation~(\ref{manypevent}) is, in general, more involved than one described by equation~(\ref{intuitiveexp}). In particular, it can be nonmonotonic in $\alpha$. 
  
An exception of this general behavior is described in \cite{ra}: A bunched final event with all particles in one mode can be reached by one many-particle path only, leaving no place for any phase-dependent interference. The enhancement of such events purely relies on bosonic bunching, and not on the interference of different physical processes. Since no amplitudes with different phases appear, greater indistinguishability always leads to an enhancement of such events and to $P_{N}>P_{N-1}> \dots >P_0$, which implies a monotonic transition. Most importantly, this enhancement does not ``fortuitously turn [..] out to be monotonic'' as stated in reference~\cite{bjork}, but 
the absence of different many-particle paths forces every transition to a bunched final state of the form $(N,0,\dots, 0)$ to be monotonic.

In contrast to the hitherto established scenario, the examples put forward in reference~\cite{bjork} do not mediate a quantum-to-classical transition. The transition given in reference~\cite{bjork} interpolates from a pure state to another pure state, without loss of interference capability at any stage. 
 Let us take equation~(8) in reference~\cite{bjork} as an example,
\beq
\ket{\psi_1^D(\gamma)}= \cos{\left(\pi/4+\gamma/2\right)} \ket{1,0} + \sin{\left(\pi/4+\gamma/2\right)} \ket{0,1} ,\nonumber \\ \label{eq:eq8bj}
\eeq
where $\ket{1,0}$ ($\ket{0,1}$) denotes a horizontally (vertically) polarized single-photon state, and $\gamma$ changes from $0$ to $\pi/2$, inducing a transition of the state from $\left(\ket{1,0}+\ket{0,1}\right)/\sqrt{2}$ to $\ket{0,1}$. The state is projected onto $\ket{\xi_1^D}= \cos{\left(\pi/8\right)} \ket{1,0} - \sin{\left(\pi/8\right)} \ket{0,1}$, resulting in the projection probability $\left|\braket{\xi_1^D}{\psi_1^D(\gamma)}\right|^2$, which is nonmonotonic as a function of $\gamma$. However, this transition cannot be identified as a quantum-to-classical transition by any means; it neither includes a transition to a mixed state nor does it involve any which-path information that degrades interference capability. The nonmonotonicity exhibited by the other examples in reference~\cite{bjork} (figures 3 and 5) has the same origin, none of  these examples is related to the quantum-to-classical transition.

In reference~\cite{bjork}, $\gamma$  is claimed to vary ``distinguishability", but that ``distinguishability" is used in a different sense compared with the generally accepted usage. Distinguishability is generally used to indicate the distinguishability of different processes (e.g. different paths of particles), i.e. distinguishability degrades interference~\cite{englert,hong,kwiat,Bocquillon}. On the other hand, the ``distinguishability" in reference~\cite{bjork} affects the overlap of the state $|\psi_1^D(\gamma)\rangle$ with an unbiased (in terms of its basis $\left\{\ket{1,0},\ket{0,1}\right\}$) state $|\psi_1^D(0)\rangle=\left(\ket{1,0}+\ket{0,1}\right)/\sqrt{2}$, i.e. it governs the magnitude of $\left|\langle\psi_1^D(0)|\psi_1^D(\gamma)\rangle\right|^2$. 
Therefore, while general distinguishability adjusts the degree to which probabilities instead of probability amplitudes are added, the ``distinguishability" in reference~\cite{bjork} changes the very probabilities themselves: In equation~(\ref{eq:eq8bj}), $\gamma$ defines the probability to populate the horizontal and the vertical modes.
Formally, $\gamma$ varies the \emph{predictability}~\cite{englert}, i.e.~the capability to predict in which mode the particle resides, rather than the degree of interference between different modes. As a result, the ``distinguishability" in reference~\cite{bjork} does not mediate the quantum-to-classical transition. If the degree of interference is varied, a monotonic rather than a nonmonotonic probability transition emerges, evident from equation (14) of reference~\cite{bjork}.

\begin{figure}[t]
\raggedleft{\includegraphics[width=\linewidth]{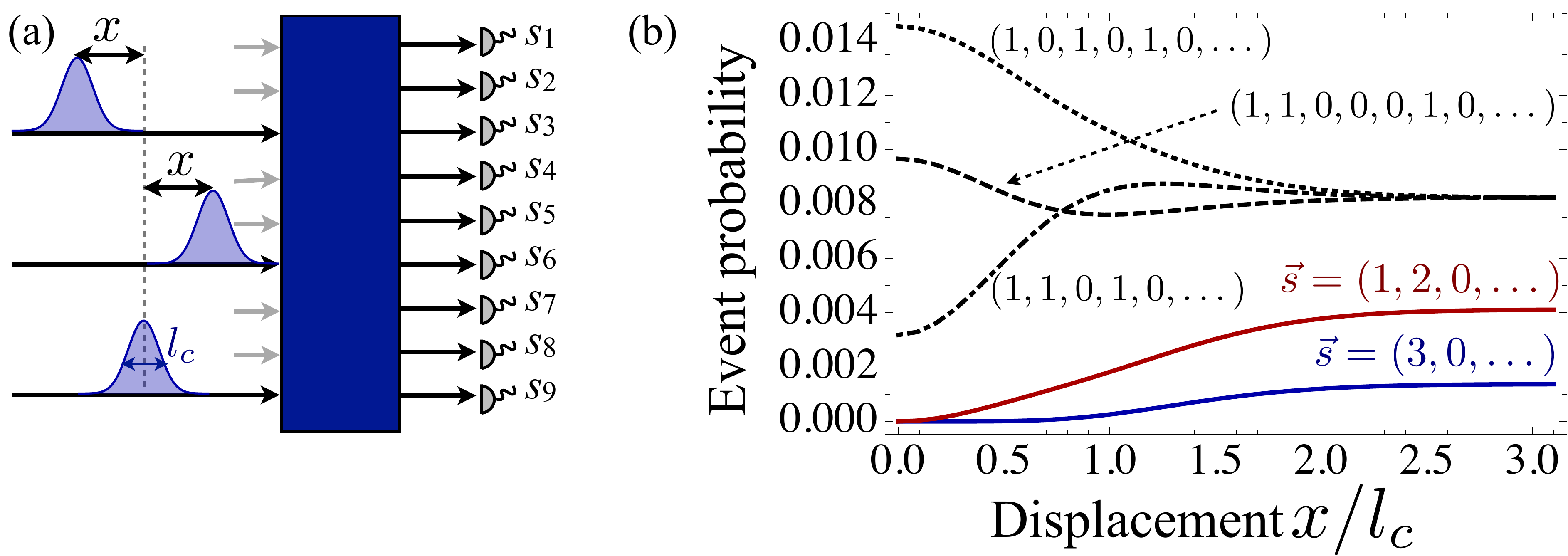}}
\caption{Nonmonotonicity in three-fermion interference. (a) Fermions of coherence length $l_c$ are prepared in the state $\vec r=(0,0,1,0,0,1,0,0,1)$,  i.e.~only the third, the sixth, and the last input modes are populated.  
The particles are delayed by a displacement $x$ with respect to each other and scatter off a 9-mode Fourier-multiport, for which every single-particle probability is $1/9$ \cite{NJPManyPInterference}. The number of particles in each output mode, represented as $\vec s=(s_1,s_2,s_3,\dots)$, is measured. (b) Probabilities for different final events. For $x \gg l_c$, the particles can be treated as distinguishable, and the event probabilities can be obtained by combinatorics, i.e.~$P(\vec s=(1,1,1,\dots))= 3!/9^3,~P(\vec s=(2,1,\dots))= 2!/9^3,~P(\vec s=(3,\dots))= 1/9^3$. When the displacement is reduced, the quantum realm is attained. The event probabilities for $\vec s=(1,2,\dots)$ (red solid) and $\vec s=(3,\dots)$ (blue solid) decay monotonically with increasing interference capability and eventually vanish for $x=0$. The events allowed by the Pauli principle (black dashed, dotted and dash-dotted lines)  can evolve in a nonmonotonic fashion, in close analogy to the effects observed in \cite{ra}. The event probabilities are computed in direct analogy to the methods developed in \cite{ra,tichy}, taking into account the anti-commutativity of fermionic creation operators. }
\label{fig:fermions}
\end{figure}

In contrast to the statements regarding fermions in reference~\cite{bjork}, many fermions can interfere collectively in a non-trivial way beyond Pauli exclusion effects: As soon as the number of available modes is larger than the number of particles, fermions can collectively interfere just as bosons do \cite{NJPManyPInterference,TutorialMalte}. In particular, many-fermion interference is governed by the very same principles as many-boson interference \cite{NJPManyPInterference}. As a consequence, many-fermion signals can exhibit nonmonotonic features just like many-boson signals, as evident from figure~\ref{fig:fermions},  which shows the  probabilities of exemplary events for three interfering fermions of varying distinguishability in a 9-mode Fourier-multiport. The experimental observation of such fermionic nonmonotonicity is admittedly challenging, but recent developments in the simulation of fermions via entangled photons \cite{Sansoni} or the interference of electrons \cite{Bocquillon} feed the hope that such experiments become feasible in future.

In conclusion, we have shown that the criticism in reference~\cite{bjork} on references~\cite{ra,tichy} is inappropriate. In the first place, references~\cite{ra,tichy} do not claim that a nonmonotonically evolving probability should always be regarded as  a signature for the quantum-to-classical transition.  Instead, references~\cite{ra,tichy} report that, in the multi-particle quantum-to-classical transition, probabilities are typically nonmonotonic, which we generalized here for fermions in figure~\ref{fig:fermions}.  The  examples presented in reference~\cite{bjork} cannot be regarded as quantum-to-classical transitions by any means, and the nonmonotonic probabilities in reference~\cite{bjork} are not rooted in the quantum-to-classical transition, but in a unitary evolution of pure quantum states.

We thank Gunnar Bj\"{o}rk and Saroosh Shabbir for usuful discussions. This work was supported in part by the National Research Foundation of Korea (Grant No.~2013R1A2A1A01006029 and 2011-0021452). H.-T.L. acknowledges support from the National Junior Research Fellowship (Grant No.~2012-000642). A.B. acknowledges partial support through the COST Action MP1006 ``Fundamental Problems in Quantum Physics", and by DFG. M.C.T. would like to thank the Danish Council for Independent Research.


\vspace{0.3cm}

\begin{thebibliography}{10}

\bibitem{bjork}
Bj\"{o}rk G and Shabbir S 2014 \emph{New J. Phys.} \textbf{16} 013006, also available in 2013 arXiv:1312.0055

\bibitem{ra}
Ra Y-S, Tichy M C, Lim H-T, Kwon O, Mintert F, Buchleitner A, and Kim Y-H 2013 \emph{Proc. Natl Acad. Sci. USA} \textbf{110} 1227--1231

\bibitem{tichy}
Tichy M C, Lim H-T, Ra Y-S, Mintert F, Kim Y-H, and Buchleitner A 2011 \emph{Phys. Rev. A} \textbf{83} 062111

\bibitem{TutorialMalte}
Tichy M C 2014 \emph{J. Phys. B: At. Mol. Opt. Phys.} \textbf{47} 103001

\bibitem{Laloe}
Mullin W J and Lalo\"e F 2008 \emph{Phys. Rev. A} \textbf{78}  061605 

\bibitem{Feynman}
Feynman R P, Leighton R B, and Sands M 1965 \emph{The Feynman Lectures on Physics} Vol. \textbf{III}  Addison Wesley

\bibitem{arndt1}
Hornberger K, Uttenthaler S, Brezger B, Hackerm\"{u}ller L, Arndt M, and Zeilinger A 2003 \emph{Phys. Rev. Lett.} \textbf{90} 160401

\bibitem{arndt2}
Hackerm\"{u}ller L, Hornberger K, Brezger B, Zeilinger A, and Arndt M 2004 \emph{Nature} \textbf{427} 711-714

\bibitem{englert}
Englert B-G 1996 \emph{Phys. Rev. Lett.} \textbf{77} 2154--2157

\bibitem{hong}
Hong C K, Ou Z Y and Mandel L 1987 \emph{Phys. Rev. Lett.} \textbf{59} 2044--2046

\bibitem{kwiat}
Kwiat P G, Steinberg A M, and Chiao R Y 1992 \emph{Phys. Rev. A} \textbf{45} 7729--7739

\bibitem{Bocquillon}
Bocquillon E \textit{et al.} 2013 \emph{Science} \textbf{339} 1054--1057

\bibitem{NJPManyPInterference} 
Tichy M C, Tiersch M, Mintert F and Buchleitner A 2012 \emph{New J. Phys.} \textbf{14} 093015

\bibitem{Sansoni}
Sansoni L \textit{et al.} 2012 \emph{Phys. Rev. Lett.} \textbf{108} 010502

\end{thebibliography}
\end{document}